\documentclass[twocolumn,10pt]{aastex631}

\usepackage[utf8]{inputenc}
\usepackage{CJK}
\usepackage[utf8]{inputenc}
\usepackage[flushleft]{threeparttable}
\usepackage{longtable}
\usepackage{multirow}
\usepackage[T1]{fontenc}
\usepackage{bm}
\usepackage{enumitem}
\usepackage{array}
\usepackage{mathtools}
\makeatletter

\newcommand{\Rmnum}[1]{\expandafter\@slowromancap\romannumeral #1@}
\makeatother

\shorttitle{Explosion mechanism of SNe II}
\shortauthors{Fang, Nagakura \& Moriya}

\begin{document}


\title{Reconciling tension between neutrino-driven explosion mechanism and optical light curve modeling of type II supernovae}

\title{Reconciling the Tension Between Light Curve Modeling of Type II Supernovae and Neutrino-Driven Core-Collapse Supernovae Models with Late-Phase Spectroscopy}
\author[0000-0002-1161-9592]{Qiliang Fang}\affiliation{National Astronomical Observatory of Japan, National Institutes of Natural Sciences, 2-21-1 Osawa, Mitaka, Tokyo 181-8588, Japan}

\author[0000-0002-7205-6367]{Hiroki Nagakura}\affiliation{National Astronomical Observatory of Japan, National Institutes of Natural Sciences, 2-21-1 Osawa, Mitaka, Tokyo 181-8588, Japan}

\author[0000-0003-1169-1954]{Takashi J. Moriya}
\affiliation{National Astronomical Observatory of Japan, National Institutes of Natural Sciences, 2-21-1 Osawa, Mitaka, Tokyo 181-8588, Japan}
\affiliation{Graduate Institute for Advanced Studies, SOKENDAI, 2-21-1 Osawa, Mitaka, Tokyo 181-8588, Japan}
\affiliation{School of Physics and Astronomy, Monash University, Clayton, VIC 3800, Australia}

\begin{abstract}
Type II supernovae (SNe II) are the most frequently observed outcome of core-collapse explosions and provide a valuable window into the physical mechanisms governing the deaths of massive stars. However, estimates of explosion properties based on optical light curve modeling often show tension with the predictions of modern neutrino-driven explosion models. In particular, when based on light curves from the explosions of red supergiant (RSG) tied to specific stellar wind models, many SNe II are found to originate from low-mass progenitors yet exhibit unusually high explosion energies ($E_{\rm K}$), far exceeding theoretical predictions.
In this study, we incorporate late-phase (nebular) spectroscopy to estimate the helium core mass of the progenitor ($M_{\rm He\,core}$), which serves as an additional constraint to break degeneracies in light curve modeling. This approach is applied to a sample of 32 well-observed SNe II, using a light curve model grid constructed from RSGs with arbitrarily stripped hydrogen-rich envelopes, rather than assuming a fixed wind model. Examining the resulting correlations among the physical parameters, we find that the tension between the observed $M_{\rm He\,core}$-$E_{\rm K}$ and $E_{\rm K}$-$M_{\rm Ni}$ relations and those predicted by neutrino-driven explosion models has significantly lessened by incorporating nebular spectroscopy in light curve modeling. This study highlights the crucial role of nebular spectroscopy in interpreting SNe II observations and provides support to the neutrino-driven explosion mechanism as the dominant engine powering these events.

\end{abstract}
 


\section{INTRODUCTION}
Core-collapse supernovae (CCSNe) announce the explosive death of massive stars with zero-age-main-sequence (ZAMS) mass $M_{\rm ZAMS}\,\gtrsim\,8\,M_{\odot}$. At the late stage of stellar evolution, when the fuel in the core is exhausted, nuclear fusion cannot provide sufficient pressure, and the core collapses under its own gravity, leading to the formation of a neutron star or a black hole \citep{heger03}. During this process, a strong shock is generated, which disrupts the star and ejects its outer stellar material ($ejecta$). These catastrophic events are observed as CCSNe, and they are important probes of the final stages of massive star evolution, nucleosynthesis, and the physical conditions of extreme environments.

A central unsolved problem in modern astrophysics is understanding the physical mechanism that drives the explosion. Among the leading candidates is the neutrino-driven mechanism, in which neutrino heating, aided by turbulence, revives the stalled shock and propels it outward through the stellar envelope, resulting in a successful SN explosion \citep{burrows95,takiwaki12,couch15,nakamura16,muller17,nagakura18,vartanyan18,muller19,nagakura19,vartanyan22}. This process can be amplified or damped by the magnetic field \citep{matsumoto20,matsumoto22,varma23,matsumoto24,nakamura25} or rotation \citep{marek09,suwa10,takiwaki16}. Alternative scenarios include magneto-rotation driven explosions \citep{bis18,obergaulinger20,varma21,powell23} and jittering jet \citep{soker19,soker22}, which may operate in specific progenitor configurations. The readers may refer to \citet{mezzacappa20,burrows21,muller24,yamada24,janka25} for recent reviews.

Due to the extremely short timescale and the opaque nature of the stellar collapse, directly observing the explosion engine remains a major challenge. However, explosive neucleosynthesis leaves behind observational signatures. In particular, radioactive decay of $^{56}$Ni is an important power source of late-phase SN light curve. At the same time, a fraction of the energy released by core-collapse is converted into the thermal and kinetic energy of the expanding ejecta. Both the $^{56}$Ni mass ($M_{\rm Ni}$) and the energy stored in the ejecta ($E_{\rm K}$) can be constrained from modeling the SN light curve. The distributions of these quantities, as well as their correlations with progenitor properties, provide powerful diagnostics of the observables predicted by the underlying core-collapse mechanism(s). This has been made possible thanks to recent advances in core-collapse simulations, which now allow the explosion to be followed to late times and provide predictions for observable quantities (see, e.g., \citealt{burrows24a,janka24,janka25}).

Type II SN (SNe II) are the most commonly observed subtype of CCSNe \citep{li06,perley20,sharma24}. These objects are characterized by the presence of strong hydrogen lines in their spectra, indicating that their progenitors retained a massive hydrogen-rich envelope at explosion \citep{nomoto95,filippenko97}. The light curves of SNe II often exhibit a plateau, with properties mainly determined by the hydrogen-rich envelope mass $M_{\rm Henv}$, the radius of the progenitor $R_{\rm prog}$ at shock breakout, the explosion energy $E_{\rm K}$ and the $^{56}$Ni mass $M_{\rm Ni}$ \citep{kasen09,goldberg19}. 

Although considerable efforts have been made to infer these parameters through light curve modeling, this process is hindered by strong degeneracies \citep{goldberg20}. As a result, it is often necessary to introduce additional assumptions, for example, an $M_{\rm ZAMS}$-$M_{\rm Henv}$-$R_{\rm prog}$ relation based on stellar evolution models (see, e.g., \citealt{morozova18,martinez22a,martinez22b,martinez22c,moriya23,subrayan23,das25}), in order to reduce the dimension of the parameter space and enable meaningful interpretation. 

However, this approach presents several limitations, particularly in the context of understanding the core-collapse mechanism: (1) the $M_{\rm ZAMS}$-$M_{\rm Henv}$-$R_{\rm prog}$ relation is sensitive to uncertainties in the mass-loss mechanism (e.g., binary interaction; see \citealt{ouchi17,eldridge18,fragos23,matsuoka23,ercolino24}), RSG wind model \citep{mauron11,renzo17,yang23,zapartas25}, metallicity and internal mixing \citep{dessart13,dessart14}, uncertainties in all these factor not only affect the inferred $M_{\rm Henv}$ but also introduce biases in the derived $E_{\rm K}$; and (2) the inferred $M_{\rm ZAMS}$ is effectively constrained by $M_{\rm Henv}$ and $R_{\rm prog}$, parameters that are largely decoupled from the core structure and the explosion mechanism itself.

As a result, this method, based solely on light curve modeling and progenitor models with a fixed $M_{\rm ZAMS}$-$M_{\rm Henv}$-$R_{\rm prog}$ relation, potentially leads to inconsistencies between the observed and theoretically predicted relation between the helium core mass $M_{\rm He\,core}$ (equivalent to $M_{\rm ZAMS}$ in the aforementioned works) and $E_{\rm K}$, which is a key diagnostic of the core-collapse mechanism. For example, in 2 well-observed SNe II, SNe 2023ixf and 2024ggi, their inferred $M_{\rm He\,core}$ are relatively low (corresponding to $M_{\rm ZAMS}\,\sim$\,10 to 12\,$M_{\rm \odot}$ for the adopted progenitor models), while $E_{\rm K}$ reaches 1.5-3.0\,foe\footnote{1\,foe\,=\,10$^{51}$\,erg} \citep{bersten24,singh24,moriya24,chen25}. When this method is applied to a large sample of SNe II, \citet{martinez22c} find many SNe II having low $M_{\rm He\,core}$ (corresponding to $M_{\rm ZAMS}$\,\(<\)\,12\,$M_{\rm \odot}$ for the progenitor models) but $E_{\rm K}$\,\(>\)\,1.0 foe (see also \citealt{silva24,subrayan23,das25}), far exceeding the predicted $\sim$\,0.1\,foe by standard neutrino-driven explosions in this mass range \citep{muller17,muller19,burrows24a,janka24}.

These parameter estimations can be improved by incorporating late-phase (nebular) spectroscopy into light curve modeling \citep{fang25a,fang25b,fang25c,fang25d}. Approximately one year after explosion, the ejecta becomes optically thin and enters the nebular phase, where the spectrum is dominated by forbidden emission lines. In SNe II, one of the most prominent features is the [O I] $\lambda\lambda$6300,6363 doublet, which serves as a proxy of the oxygen mass in the ejecta \citep{fransson89,maeda07,hanin15,jerkstrand15,jerkstrand17,fang19,dessart20,dessart21a,dessart21b,hiramatsu21,fang22,dessart23,2023ixf_nebular,2024ggi_nebular} and can be used to independently estimate $M_{\rm He\,core}$ or log\,$L_{\rm prog}$. This raises an interesting question of whether the apparent tension between light curve modeling and the neutrino-driven explosion mechanism can be reconciled by this additional constraint. The attempt in this paper is to extend our previous studies, where we incorporate nebular spectroscopy into light curve modeling, which are based on RSG progenitor models with partially-stripped hydrogen-rich envelope. This analysis is applied to a sample of 32 well-observed SNe II. The observed $M_{\rm He\,core}$-$E_{\rm K}$ and $E_{\rm K}$-$M_{\rm Ni}$ relations are compared with the predictions from recent multi-dimensional core-collapse supernova simulations. More specifically, we shall take the following steps in the present study:

\begin{itemize}
    \item We constructed a grid of light curves from the explosions of RSG progenitors with different initial mass $M_{\rm ZAMS}$ and hydrogen-rich envelopes artificially stripped to varying degrees. By treating $M_{\rm Henv}$ as a free parameter, we account for the large uncertainties in RSG mass-loss rates, rather than relying on fixed wind prescriptions from stellar evolution models that can bias the inferred $E_{\rm K}$. These models are used to fit observed SNe II light curves and infer the progenitor and explosion properties. For a fixed log\,$L_{\rm prog}$, there is always a combination of \{$M_{\rm Henv}$, $E_{\rm K}$, log\,$M_{\rm Ni}$\} that can reproduce the observed light curves (\citealt{fang25a,fang25b});
    \item To break the degeneracy of the inferred parameters with log\,$L_{\rm prog}$, we incorporate nebular spectroscopy as additional constraint. By estimating log\,$L_{\rm prog}$ from nebular spectroscopy, we uniquely select a progenitor log\,$L_{\rm prog}$ and the associated inferred parameters from light curve modeling (\citealt{fang25c,fang25d});
    \item Recent works have shown that log\,$L_{\rm prog}$ is tightly correlated with the helium core mass ($M_{\rm He\,core}$) and carbon-oxygen (CO) core mass as the luminosity mostly originate from the CO core before collapse (\citealt{farrell20,temaj24,schneider24}). \citet{fang25c} further confirmed that these correlations hold across several stellar evolution models, including \texttt{MESA} (progenitor models taken from \citealt{fang23,temaj24}), \texttt{KEPLER} (\citealt{kepler16,sukhbold18,ertl20}) and \texttt{HOSHI} (\citealt{takahashi18,takahashi21,takahashi23}), despite these models having different physical assumptions, such as the treatment of overshooting and the adopted nuclear reaction network. The readers may refer to these papers for more details. The universal relation allows us to use log\,$L_{\rm prog}$ as a proxy for core mass and directly investigate its relation with \{$E_{\rm K}$, log\,$M_{\rm Ni}$\} inferred from light curve modeling, providing a more physically meaningful diagnosis of the core-collapse mechanism.
\end{itemize}

This paper is organized as follows: in \S2, we describe the numerical setup, including the construction of the light curve model grid, the fitting procedure used to constrain physical parameters, and the method by which nebular spectroscopy is incorporated into the analysis. In \S3, we examine the correlations between log\,$L_{\rm prog}$ estimated from nebular spectroscopy and the explosion parameters  \{$E_{\rm K}$, log\,$M_{\rm Ni}$\} inferred from light curve modeling for the 32 SNe II in the sample and compare them with the predictions of recent core-collapse simulations. Conclusions are left to \S4.

\section{Numerical Setup and Parameter Inference}
The sample for this study is the same as that of \citet{fang25d}, comprising 32 SNe II with both multi-band plateau-phase light curves and nebular spectroscopy, which together allow for simultaneous constraints on surface and core properties. We begin by fixing the log\,$L_{\rm prog}$ of the RSG progenitor with its hydrogen-rich envelope partially-stripped to different degrees, and at each log\,$L_{\rm prog}$ value, the physical parameters \{$M_{\rm Henv}$, $E_{\rm K}$, log\,$M_{\rm Ni}$\} are inferred by fitting the plateau phase light curve. We then use nebular spectroscopy to independently estimate log\,$L_{\rm prog}$ to break the degeneracy inherent in light curve modeling. Below, we summarize this workflow, and the readers may refer to \citet{fang25a,fang25b,fang25c,fang25d} for full details.

\subsection{Light curve grid}
A grid of solar metallicity, non-rotating RSG models are evolved using \texttt{MESA} (version \texttt{r-23.05.01}; \citealt{paxton11, paxton13, paxton15, paxton18, paxton19, mesa23}). The initial mass $M_{\rm ZAMS}$ ranges from 10 to 18\,$M_{\rm \odot}$ in 1\,$M_{\rm \odot}$ step. The mixing length parameter ($\alpha_{\rm MLT}$) is set to 2.5, and the exponential overshooting parameter ($f_{\rm ov}$) is 0.004. These parameters are chosen to ensure that the progenitors reach an effective temperature $T_{\rm eff}$ of approximately 3650\,K at the end of the calculation, similar to the values of field RSGs (\citealt{levesque05,levesque06,massey16}; see discussions in \citealt{fang25d}), and follow the similar $M_{\rm ZAMS}$-$M_{\rm He\,core}$ relation as the \texttt{KEPLER} models \citep{kepler16}, which are frequently employed as the initial conditions for modeling SN II light curves \citep{morozova18,moriya23,vartanyan24}, nebular spectroscopy \citep{jerkstrand12,jerkstrand14}, and core-collapse processes \citep{sukhbold18,ertl20,patton20,sukhbold20,burrows21,burrows24a,burrows24b} that will be discussed in later sections.

The models are evolved from the pre-ZAMS stage, with the hydrogen-rich envelope artificially removed to varying degrees after the core helium-burning phase, and evolution continues until core-carbon depletion, as further evolution will not affect the plateau phase light curves \citep{eldridge19,fang25a}. Similar to the discussion in \cite{fang25d}, varying $M_{\rm ZAMS}$ should be understood as a means to generate models with different RSG radii $R_{\rm prog}$ (therefore log\,$L_{\rm prog}$ at fixed $T_{\rm eff}$ assumed in this work), as it plays a more fundamental role in light curve modeling than $M_{\rm ZAMS}$ itself. Throughout this work, although we will sometimes use the term M12 to refer to models with $M_{\rm ZAMS}$\,=\,12\,$M_{\rm \odot}$ (with similar naming conventions for other $M_{\rm ZAMS}$), it actually represents RSG models with log\,$L_{\rm prog}$\,=\,4.72 or $M_{\rm He\,core}$\,=\,3.15\,$M_{\rm \odot}$ (Table~1 of \citealt{fang25d}).

The explosion is initiated as a thermal bomb using \texttt{MESA}, with varying amounts of energy injected into the innermost region. Throughout this work, we define "explosion energy" as the energy of the thermal bomb minus the binding energy of the star. Once the shock has propagated to 0.05\,$M_{\rm \odot}$ below the stellar surface, different amounts of radioactive $^{56}$Ni are uniformly distributed within the helium core (the compact remnant is removed), and the material is smoothed using a boxcar mixing scheme (\citealt{kasen09,dessart12,dessart13,snec15,fang25a}). It should be noted that material mixing can influence the duration of the plateau \citep{utrobin07,utrobin17}, but a comprehensive discussion of this effect is beyond the scope of the present work. Accordingly, all discussions in this work are based on the boxcar mixing scheme.

The model is then passed to the public version of \texttt{STELLA} (\citealt{blinnikov98, blinnikov00, blinnikov06}) to compute the multi-band light curve with 40 frequency bins and 800 spatial zones. No circumstellar material (CSM) is introduced in the models. 
        
Roughly speaking, the ranges of $M_{\rm Henv}$, $E_{\rm K}$ and log\,$M_{\rm Ni}$ are: [3, 7-10]\,$M_{\rm \odot}$, [0.1, 1.5]\,foe and [-3, -0.8]\,dex ($M_{\rm Ni}$ in solar unit). The pre-computed grid consists of 58,846 models. The MESA inlists used to generate the progenitor and explosion models are available on Zenodo under an open-source Creative Commons Attribution 4.0 International license: doi:10.5281/zenodo.13953755.

\subsection{Fitting observed light curves}
For each fixed log\,$L_{\rm prog}$, we define the intrinsic variables $\theta$\,=\,\{$M_{\rm Henv}$, $E_{\rm K}$, log\,$M_{\rm Ni}$\}. The interpolation method similar to that described in \citet{F18} is employed to compute the multi-band light curves for arbitrary combinations of \{$M_{\rm Henv}$, $E_{\rm K}$, log\,$M_{\rm Ni}$\}, even those not among the pre-computed grid. The \texttt{Python} package \texttt{emcee} is used to infer the optimized values of \{$M_{\rm Henv}$, $E_{\rm K}$, log\,$M_{\rm Ni}$\} for each fixed log\,$L_{\rm prog}$. A degeneracy between the inferred parameters and log\,$L_{\rm prog}$ is evident, as shown in the left panel of Figure~\ref{fig:weight}: there is no significant difference in the best-fit light curves across models from M10 to M18, and the inferred values of \{$M_{\rm Henv}$, $E_{\rm K}$, log\,$M_{\rm Ni}$\} depend on log\,$L_{\rm prog}$. In the next steps, we break this degeneracy by incorporating additional constraints from nebular spectroscopy.

\subsection{Inferring log\,$L_{\rm prog}$ from nebular spectroscopy}
For each individual SNe, we calculate the fractional flux of [O I], $f_{\rm [O I]}$, and its regulated form $f_{\rm [O I],reg}$, defined as fractional flux of [O I] excluding H$\alpha$, after the local galaxy flux is removed (see also \citealt{liu25}). These values are employed to estimate $M_{\rm ZAMS}$ by comparing with the nebular spectroscopy models (\citealt{jerkstrand12,jerkstrand14}). Specially, $M_{\rm ZAMS}$ estimated in this way is denoted as $M_{\rm ZAMS,neb}$. Note that $M_{\rm ZAMS,neb}$ should be understood as a proxy of oxygen mass in the ejecta.

In our sample, 13 SNe II have both nebular spectroscopy and pre-SN images (golden sample), allowing for the direct comparison of $M_{\rm ZAMS,neb}$ and log\,$L_{\rm prog}$ \footnote{The log\,$L_{\rm prog}$ values and the uncertainties are taken from \citet{DB18,rui19_17eaw,2017eaw,kilpatrick23,vandyk24}; but see \citet{beasor25} that the uncertainties may be underestimated.}. As shown in Figure~8 of \cite{fang25c}, a strong correlation is discerned (with SN 2013ej as an outlier). This relation is referred to as empirical mass-luminosity relation (MLR), and is used to estimate log\,$L_{\rm prog}$ of other objects: $M_{\rm ZAMS,sub}$ and log\,$L_{\rm prog}$ of the objects in the golden sample are randomized according to their uncertainties, and a linear regression is performed on this sub-sample. For all other objects, we generate a random sample of $M_{\rm ZAMS,neb}$, and estimate log\,$L_{\rm prog}$ according to this linear regression. The above step is repeated 10,000 times. This process yields a distribution of log\,$L_{\rm prog}$, denoted as $P$(log\,$L_{\rm prog}$), for each SNe II in the sample based on its nebular spectroscopy (Figure~\ref{fig:weight}). 
    
\subsection{Weighting the parameters with log\,$L_{\rm prog}$}
Finally, we employ the log\,$L_{\rm prog}$ converted from nebular spectroscopy to break the degeneracy in plateau phase light curve modeling as
    \[
    P(x)\,=\,\int\int P(x|{\rm log}\,L_{\rm prog})\,P({\rm log}\,L_{\rm prog})d\,{\rm log}\,L_{\rm prog}\,dx.
\]
Here, $x$ denotes either $E_{\rm K}$ or log\,$M_{\rm Ni}$; $P(x)$ is the marginal posterior distribution of $x$, while $P(x|{\rm log}\,L_{\rm prog})$ represents the posterior probability density of $x$ at a fixed log\,$L_{\rm prog}$, as obtained from light curve modeling. In practice, $P(x)$ is estimated as follows:
    \begin{enumerate}
        \item For each individual SNe, a random value \( l \) is drawn from the posterior \(\log L_{\rm prog}\) distribution established from its nebular spectroscopy. Since $l$ may not match exactly with the log\,$L_{\rm prog}$ values of the pre-computed RSG models, the corresponding $E_{\rm K}$ (log\,$M_{\rm Ni}$) is estimated via linear interpolation. 
        
        We find the index \( i \) such that  
        \[
        \log L_{{\rm prog},i} \,< \,l \,< \,\log L_{{\rm prog},i+1}.
        \]
        For example, if $l\,$=\,4.85, then we find $i$\,=\,3, and log$\,L_{{\rm prog},i}$ and log$\,L_{{\rm prog},i+1}$ (see Table~1 of \citealt{fang25d}).
        \item A value \( e_i \) is randomly drawn from the posterior \( E_{\rm K} \) distribution inferred from light curve modeling at fixed \( {\rm log}\,L_{{\rm prog},i}  \).
        \item Similarly, we obtain \( e_{i+1} \) from the posterior distribution of \( E_{\rm K} \) at fixed \( {\rm log}\,L_{{\rm prog},i+1}  \).
        \item The explosion energy $e$ corresponding to \( l \) is then estimated via linear interpolation. In cases that $l$\,\(<\)\,4.52\,dex (the minimum log\,$L_{\rm prog}$ of the progenitor model grid), $e$ is estimated from the linear extrapolation.
        \item Similar process is also conducted for log\,$M_{\rm Ni}$ distributions.
        \item The above processes are repeated for 10,000 times to construct the log\,$L_{\rm prog}$-weighted posterior distributions of $E_{\rm K}$ and log\,$M_{\rm Ni}$.
    \end{enumerate}

\begin{figure*}
\epsscale{1.}
\plotone{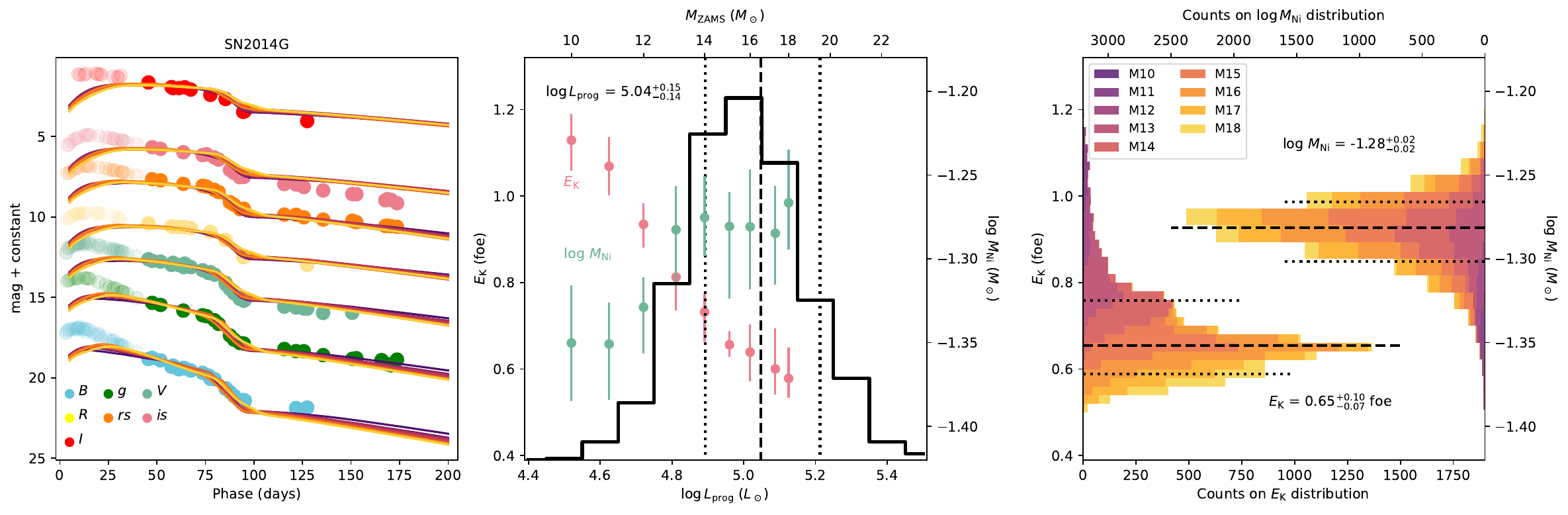}
\centering
\caption{Left panel: The optimized multi-band light curves of SN 2014G at different fixed log\,$L_{\rm prog}$, which are distinguished by colors. Middle panel: The median values and 68\% CIs of the inferred $E_{\rm K}$ (pink scatter points) and log\,$M_{\rm Ni}$ (green scatter points) from light curve modeling as function of fixed log\,$L_{\rm prog}$. The black histogram represents the log\,$L_{\rm prog}$ distribution inferred from nebular spectroscopy. Right panel: The log\,$L_{\rm prog}$-weighted distributions of $E_{\rm K}$ (represented by the left y-axis) and log\,$M_{\rm Ni}$ (represented by the right y-axis). The contributions from $E_{\rm K}$ (log\,$M_{\rm Ni}$) distributions, with different fixed log\,$L_{\rm prog}$, are distinguished by colors. The black dashed lines represent the median values of the distributions, and the black dotted lines represent the 16th and 84th percentiles.}
\label{fig:weight}
\end{figure*}

In the right panel of Figure~\ref{fig:weight}, we show the log\,$L_{\rm prog}$-weighted posterior distributions of $E_{\rm K}$ and log\,$M_{\rm Ni}$ for SN 2014G, constructed using the procedures outlined above as examples. In the middle panel, the pink and green scatter points represent the median values of the $E_{\rm K}$ (left y-axis) and log\,$L_{\rm prog}$ (right y-axis) posterior distributions, respectively, inferred from plateau light curve modeling at fixed log\,$L_{\rm prog}$, and the error bars represent the 68\% confidence intervals (CI; defined by the 16th and 84th percentile) of these distributions. The log\,$L_{\rm prog}$ distribution derived from nebular spectroscopy is shown as the black histogram. The right panel presents the log\,$L_{\rm prog}$-weighted $E_{\rm K}$ (left y-axis) and log\,$M_{\rm Ni}$ (right y-axis) distributions used in the subsequent analysis, with the relative contributions from the $E_{\rm K}$ and log\,$M_{\rm Ni}$ posteriors, each inferred from light curve modeling at different fixed log\,$L_{\rm prog}$ values, distinguished by colors.

In addition to the log\,$L_{\rm prog}$-weighting method described above, we also conduct a comparative study: we adopt the common approach to model the plateau phase light curves, where $M_{\rm Henv}$ is not treated as a free parameter as in this work; instead, it is derived from massive star models evolved with standard stellar wind prescriptions. These models impose a specific log\,$L_{\rm prog}$-$R_{\rm prog}$-$M_{\rm Henv}$ relation, enabling the inference of these parameters directly from light curve modeling without additional observational constraints such as nebular spectroscopy. We employ progenitor models that follow the similar log\,$L_{\rm prog}$-$R_{\rm prog}$-$M_{\rm Henv}$ relation as the \texttt{KEPLER} models (\citealt{kepler16}) to infer \{log\,$L_{\rm prog}$, $E_{\rm K}$, log\,$M_{\rm Ni}$\}, and investigate how the correlations in \S3 are affected by the different modeling methods. These models are referred to as 'wind models' for convenience throughout this work. The readers may refer to \S5.1 of \citet{fang25d} for details of this method.

\section{Correlations}
In this section, we compare the mutual relations between $M_{\rm He\,core}$-$E_{\rm K}$-$M_{\rm Ni}$ from observation with the predictions emerged from modern neutrino-driven explosion models, including the axis-symmetric models of \citet{bruenn16} and the 3-dimensional simulations of \citet{muller17,muller19}, \citet{burrows24a} and \citet{janka24}.

\subsection{Relation between core mass and $E_{\rm K}$}
\begin{figure*}
\epsscale{1.}
\plotone{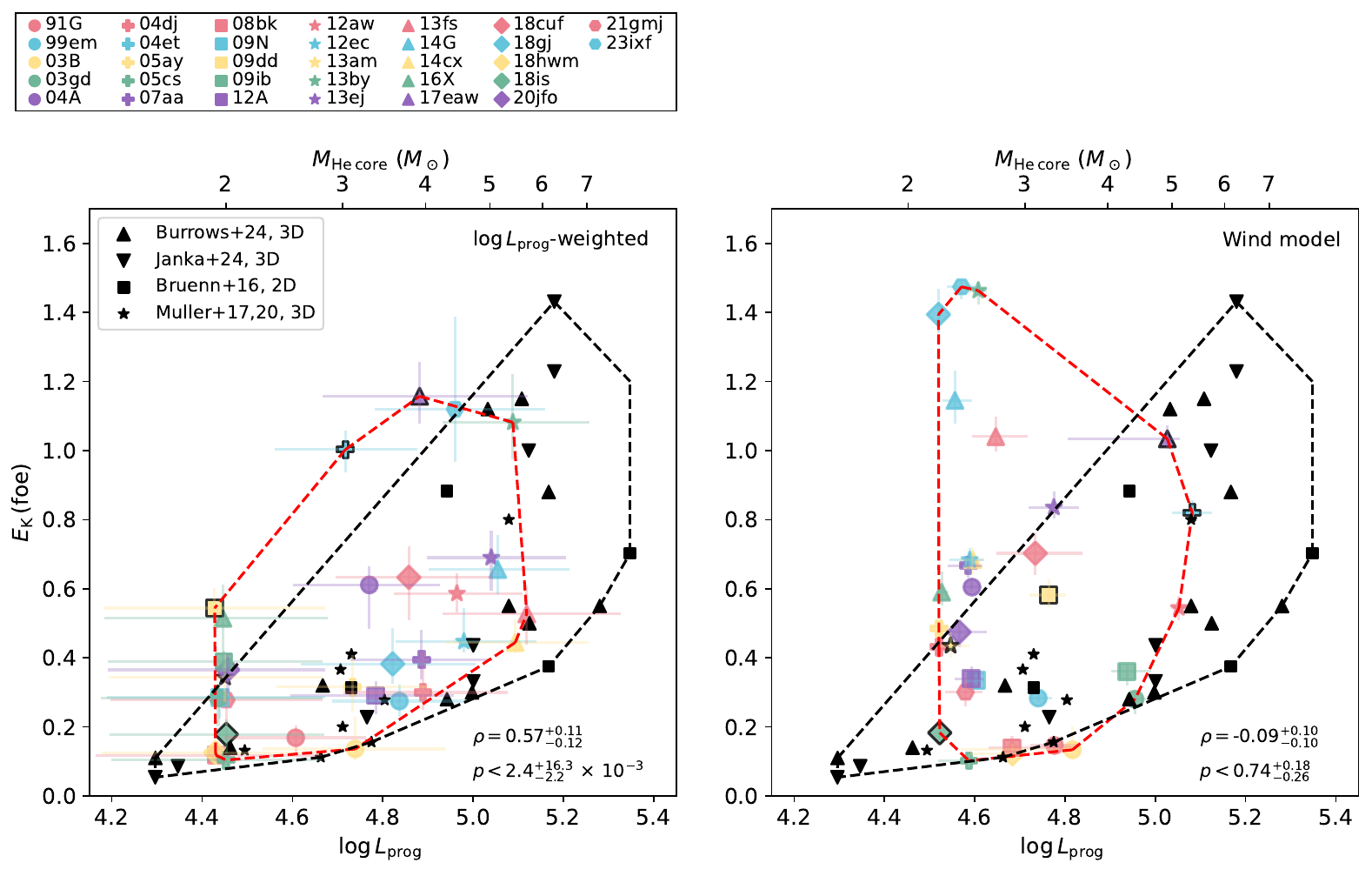}
\centering
\caption{Comparison between log\,$L_{\rm prog}$ and $E_{\rm K}$ for SNe II in the sample. Left panel: Comparison between log\,$L_{\rm prog}$ converted from nebular spectroscopy with the log\,$L_{\rm prog}$-weighted $E_{\rm K}$; Right panel: Comparison between log\,$L_{\rm prog}$ and $E_{\rm K}$ inferred from light curve modeling using wind models. Individual objects are distinguished by colors and markers. In both panels, the black scatter points are the results of core-collapse simulations. The regions surrounded by the black and red dashed lines represent the range of the models and observations respectively.}
\label{fig:mass_e}
\end{figure*}
In Figure~\ref{fig:mass_e}, we compare the progenitor luminosity log\,$L_{\rm prog}$ with the explosion energy $E_{\rm K}$ for SNe II in the sample. Individual objects are distinguished by colors and markers. The mean $E_{\rm K}$ of the full sample is 0.46\,foe. As demonstrated by \citet{fang25c}, log\,$L_{\rm prog}$ is tightly correlated with both the helium core mass and the carbon-oxygen core mass, and this correlation is largely insensitive to mixing processes and the specific details of stellar evolution models, as most of the luminosity near the onset of core-collapse comes from the CO core (\citealt{schneider24,temaj24}). Consequently, the comparison shown in Figure~\ref{fig:mass_e} reflects the relationship between $M_{\rm He\,core}$ and $E_{\rm K}$. 

The predictions of the core-collapse simulations are plotted for comparison. The progenitors' $M_{\rm He\,core}$ are converted into log\,$L_{\rm prog}$ using the following relation established in \citet{fang25c}:
\[
{\rm log}\,\frac{L}{L_{\rm \odot}}\,=\,1.47\,\times\,{\rm log}\,\frac{M_{\rm He\,core}}{M_{\rm \odot}}\,+\,4.01.
\]
Throughout this work, we treat $M_{\rm He\,core}$ and log\,$L_{\rm prog}$ as interchangeable quantities, connected via this relation.

Given the same log\,$L_{\rm prog}$, both observations and models exhibit a scatter in $E_{\rm K}$. To approximate the extent of this scatter, we use the \texttt{Python} package \texttt{scipy.spatial.ConvexHull} to compute the smallest convex set (i.e., convex hull) that encloses all data points. In Figure~\ref{fig:mass_e}, the convex hulls for the core-collapse simulations and the observational data are shown by the black and red dashed lines, respectively. These convex hulls serve as a rough estimates of the parameter spaces spanned by the models and observations.

In the left panel of Figure~\ref{fig:mass_e}, a trend that SNe II with more massive helium core (larger log\,$L_{\rm prog}$) tend to have larger $E_{\rm K}$ can be discerned. This correlation can be understood qualitatively: in the core mass range explored in this work, the compactness of the core, $\xi_{2.5}$\footnote{The compactness $\xi_{M}$ is defined as ($M/M_{\rm \odot}$)/($R(M)/1000\,$km); see \citet{oconnor}.},  increases with core mass (see, e.g., \citealt{schneider21,schneider24,schneider25} for recent works), and $\xi_{2.5}$ is demonstrated to correlate with $E_{\rm K}$ in successful neutrino-driven explosions (see, e.g., \citealt{nakamura15,muller16,burrows24a}).

To quantify the observed correlation in log\,$L_{\rm prog}$ and $E_{\rm K}$, we perform a Monte Carlo simulation. In each trial, for each SNe II, a log\,$L_{\rm prog}$ value is randomly sampled from $P$(log\,$L_{\rm prog}$), and an $E_{\rm K}$ value is randomly sampled from the log\,$L_{\rm prog}$-weighted $E_{\rm K}$ distribution (Figure~\ref{fig:weight}). We then compute the Spearman's rank correlation coefficient $\rho$ and the associated $p$-value, which quantifies the probability that such a correlation arises by chance. Repeating this process 10,000 times, we find $\rho$\,=\,0.57$^{+0.11}_{-0.12}$ with $p$\,\( < \) 2.4$^{+16.3}_{-2.2}\,\times\,10^{-3}$, indicating a statistically significant, moderately strong correlation\footnote{Throughout this work, we follow the convention of \citet{evans96} to define the strength and significance of correlations.}.

In comparison with the simulation results, the convex hulls exhibit a similar overall trend. The largest discrepancy appears at log\,$L_{\rm prog}$\,\(<\)\,4.6, where all the simulations consistently predict $E_{\rm K}\sim$\,0.1\,foe, while some SNe II show $E_{\rm K}\sim$\,0.5\,foe. This discrepancy can be explained by several observational uncertainties. 

The first and most important one is the uncertain in log\,$L_{\rm prog}$ estimated from nebular spectroscopy. As noted in \citet{fang25c}, converting $M_{\rm ZAMS,neb}$ to log\,$L_{\rm prog}$ is particularly uncertain in low luminosity regime. Such large uncertainty is also reflected in the large error bars of log\,$L_{\rm prog}$ for these objects. If we consider the most extreme scenario, i.e., shifting the log\,$L_{\rm prog}$ values of the objects showing exceed $E_{\rm K}$ (e.g., SNe 2009N, 2009ib, 2013am, 2020jfo, 2021gmj) to 4.7 ($M_{\rm ZAMS}\,\sim$\,12\,$M_{\rm\odot}$ for \texttt{KEPLER} models), the agreement between models and observations improves significantly.

Another source of uncertainty is the extinction of the host environment $E(B-V)_{\rm host}$, which directly affects the plateau luminosity and therefore the inferred  $E_{\rm K}$ in light curve modeling. In the literature, $E(B-V)_{\rm host}$ is typically estimated either from the Na ID equivalent width EW(Na ID), using the empirical relation from \citet{turatto03}, or by assuming an intrinsic color. However, both approaches have substantial limitations. (1) The EW(Na ID)-$E(B-V)$ relation has large intrinsic scatter, and the very existence of a correlation especially in large EW(Na ID) regime is debated. (2) \citet{deJaeger18} investigated a sample of SNe II with minimal host extinction, identified by the absence of detectable Na ID absorption, and showed that there is no universal intrinsic color for SNe II. In particular, the color index \(B-V\) can vary by up to \(\pm\,0.20\)\,mag at the same phase. In our sample, the 84th percentile of the total extinction $E(B\,-\,V)$ is 0.20 mag, and objects exceeding this threshold (SNe 2004et, 2009dd, 2013am, 2017eaw and 2018is) are outlined by the black edges.

For SNe II with log\,$L_{\rm prog}$\,\(>\)\,4.6, the parameter ranges and overall trends agree well with the simulation results. There are 2 outliers (SNe 2004et, 2017eaw) with high $E_{\rm K}$ that fall outside the convex hull of the models. However, both objects have $E(B-V)$ larger than the 84th percentile of the sample. Larger $E(B-V)$ values yield brighter plateau light curves and hence higher inferred explosion energies. Thus, the elevated $E_{\rm K}$ for these two objects may be an artifact of their high assumed extinction. The systematic effect of $E(B-V)$ on parameter inference is discussed in Appendix A.

In \citet{muller17}, they study the explosion of a progenitor model with $M_{\rm ZAMS}$\,=\,18\,$M_{\rm \odot}$ ($M_{\rm He\,core}$\,=\,5.3\,$M_{\rm \odot}$ and log\,$L_{\rm prog}$\,=\,5.07) and find $E_{\rm K}$\,$\sim$\,0.77\,foe. Although they suggest that this explosion energy is below average compared to observations (\citealt{utrobin05,poznanski13,chugai14,pejcha15a}), this work argues otherwise: it matches the trend seen in SNe 2013ej and 2014G and exceeds the average $E_{\rm K}$\,$\sim$\,0.46\,foe of the sample.

In the right panel, we compare log\,$L_{\rm prog}$ and $E_{\rm K}$ inferred from modeling the light curves with the wind models, and no correlation between these two quantities can be discerned. The above Monte Carlo process gives $\rho$\,=\,$-0.09^{+0.10}_{-0.10}$ with $p$\,\( < \) 0.74$^{+0.18}_{-0.26}$. Further, compared to the simulation results, the log\,$L_{\rm prog}$–$E_{\rm K}$ relation inferred from the wind models exhibits large discrepancies, as reflected in both the parameter ranges and the overall trends of the convex hulls. 

Notably, the wind models suggest some SNe II originate from relatively low mass progenitor (log\,$L_{\rm prog}\sim4.60$, corresponding to progenitor models with $M_{\rm ZAMS}\sim11\,M_{\rm \odot}$) with $E_{\rm K}$\,\(>\)\,1.2\,foe. This is not uncommon when wind models are applied: in \citet{martinez22c}, $M_{\rm ZAMS}$ (equivalent to log\,$L_{\rm prog}$) is compared with $E_{\rm K}$, and a moderately weak correlation with marginal statistical significance is discerned ($\rho\,=\,$0.34\,$\pm$\,0.10, $p\,$\(<\)\,0.08). In their Figure~8, they also show several objects with low $M_{\rm He\,core}$ (corresponding to $M_{\rm ZAMS}\,\sim$\,10\,$M_{\rm \odot}$ for their progenitor models) but $E_{\rm K}$\,\(>\)\,1.0\,foe.

The inconsistency between the log\,$L_{\rm prog}$-weighted $E_{\rm K}$ and the values inferred from wind models is addressed in \citet{fang25d} (\S5.1 therein); if an SNe II originates from massive progenitor with low $M_{\rm Henv}$, imposing a specific log\,$L_{\rm prog}$-$M_{\rm Henv}$ relation from normal wind models can lead to an overestimate of $E_{\rm K}$ but an underestimate of log\,$L_{\rm prog}$, which explains the scatter points in the upper left part in the right panel of Figure~\ref{fig:mass_e} (see also Appendix B).

\subsection{Relation between log\,$E_{\rm K}$ and log\,$M_{\rm Ni}$}
\begin{figure*}
\epsscale{1.}
\plotone{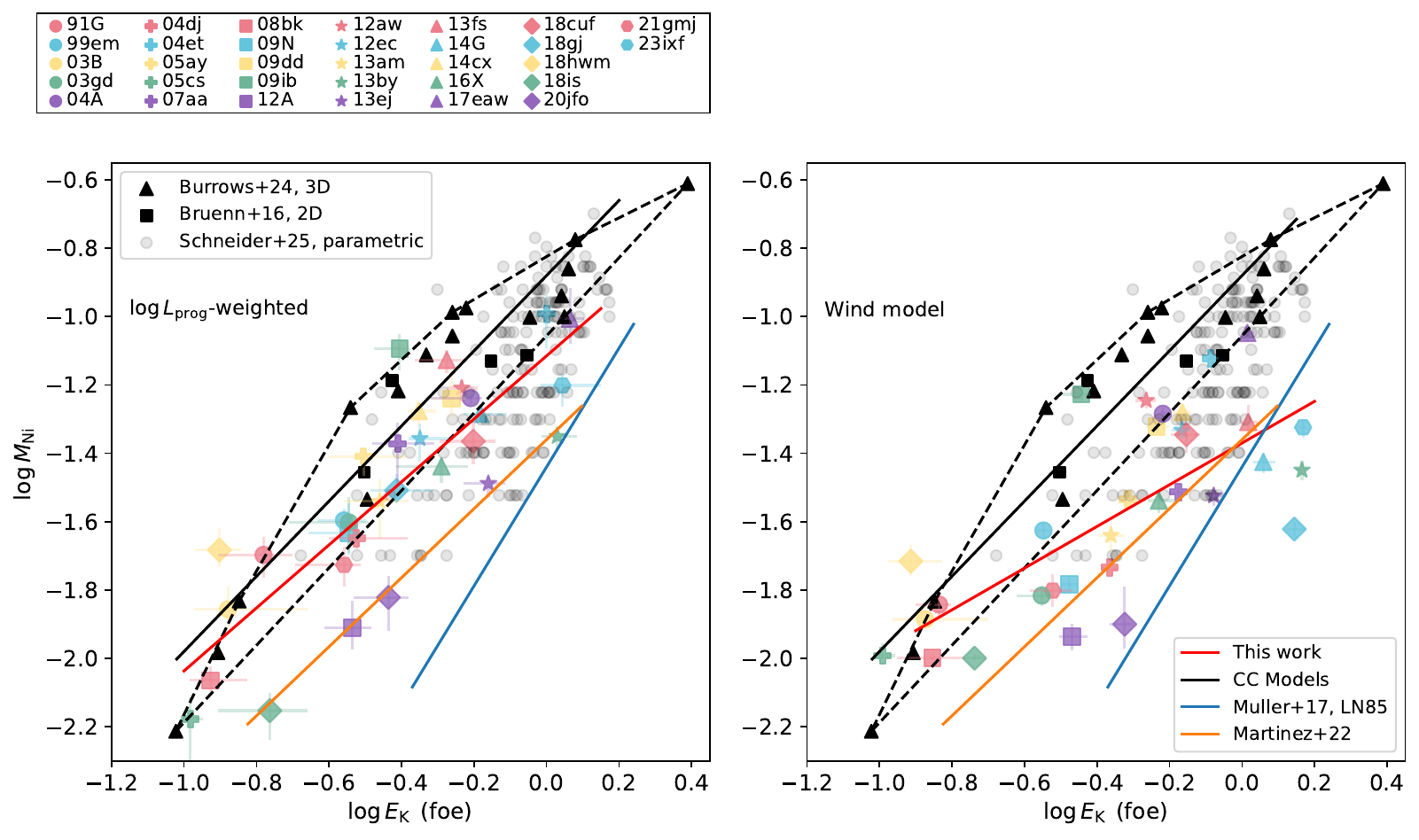}
\centering
\caption{Comparison between log\,$E_{\rm K}$ and log\,$M_{\rm Ni}$ for SNe II in the sample. Left panel: Comparison between log\,$L_{\rm prog}$-weighted log\,$E_{\rm K}$ and log\,$M_{\rm Ni}$; Right panel: Comparison between log\,$E_{\rm K}$ and log\,$M_{\rm Ni}$ inferred from light curve modeling using wind models. Individual objects are distinguished by colors and markers. In both panels, the solid black points are the results of core-collapse simulations, and the region surrounded by the black dashed lines represents a rough estimation on their range. The transparent black points are the parametric SN models in \citet{schneider25}. The log\,$E_{\rm K}$-log\,$M_{\rm Ni}$ relations from \citet{muller_t17} (blue) and \citet{martinez22c} (orange) are also plotted for comparison.}
\label{fig:e_ni}
\end{figure*}

Following core-collapse, a substantial amount of explosion energy is deposited into the dense inner regions of the ejecta, triggering explosive nucleosynthesis. One of the most important products is $^{56}$Ni, whose radioactive decay chain releases $\gamma$-ray photons and positrons that illuminate the nebular-phase emission of SNe II. The amount of synthesized $^{56}$Ni can be constrained observationally and provides a key diagnostic of the explosion mechanism. Notably, $^{56}$Ni mass, and its correlation with $E_{\rm K}$ that sets the physical environment where explosive nucleosynthesis takes place, constitutes one of the most robust, observationally testable predictions of core-collapse simulations (e.g., \citealt{sawada19,suwa19,wang24,imasheva25}).

In Figure~\ref{fig:e_ni}, we compare the inferred log\,$E_{\rm K}$ and log\,$M_{\rm Ni}$. In the left panel, these two quantities are log\,$L_{\rm prog}$-weighted, where a tight and statistical significant correlation can be discerned ($\rho\,=\,0.80^{+0.03}_{-0.04}$, $p$\,\(<\)\,10$^{-7}$). The linear regression returns
\begin{equation}
\label{eq:ek_ni_obs}
    {\rm log}\,\frac{M_{\rm Ni}}{M_{\rm \odot}}\,=\,0.94\,\times\,{\rm log}\frac{E_{\rm K}}{\rm foe}\,-1.10,
\end{equation}
as represented by the red solid line, which is broadly consistent with model prediction:
\begin{equation}
\label{eq:ek_ni_model}
    {\rm log}\,\frac{M_{\rm Ni}}{M_{\rm \odot}}\,=\,1.10\,\times\,{\rm log}\frac{E_{\rm K}}{\rm foe}\,-0.88,
\end{equation}
as represented by the black solid line. Notably, more than 80\% (26/32) of the objects in the sample fall within the convex hull of the models within uncertainties. 

The log\,$E_{\rm K}$-log\,$M_{\rm Ni}$ relations of the models from \citet{schneider25} are also plotted for comparison. The progenitors are evolved from the accretors of binary systems or the products of stellar mergers, and their core structures are used to estimate the explosive outcomes, such as $E_{\rm K}$ and $M_{\rm Ni}$, via the parametric SN model of \citet{muller16} calibrated by \citet{schneider21}. Although simplified compared to multi-dimensional core-collapse simulations, these models enable rapid estimation of explosion parameters across a wide range of core properties. As shown in the left panel of Figure~\ref{fig:e_ni}, if $E_{\rm K}$ is limited to less than 1.5\,foe, the models demonstrate a similar slope and scatter level in the log\,$E_{\rm K}$-log\,$M_{\rm Ni}$ relation comparable to the observations. We speculate that the core properties of SNe II likely display a diversity similar to that of the progenitor models in \citet{schneider25}. However, it should be noted that the parametric SN models may oversimplify explosive burning, and accurate estimation on the electron fraction $Y_{\rm e}$, which is crucial for iron-group nucleosynthesis, requires sophisticated treatments of multi-group neutrino transport (see, e.g., \citealt{wanajo23,fischer24} for recent reviews). Further work is needed to assess the true level of scatter in the log\,$E_{\rm K}$-log\,$M_{\rm Ni}$ relation that may arise from the diversity of progenitor core structures in \citet{schneider25}.

In the right panel, we compare log\,$E_{\rm K}$ and log\,$M_{\rm Ni}$ inferred from wind models. Although there is also a statistically significant strong correlation ($\rho\,=\,0.65^{+0.03}_{-0.03}$, $p$\,\(<\)\,1.7$^{+2.1}_{-1.1}\,\times 10^{-4}$), the linear regression returns
\begin{equation}
    {\rm log}\,\frac{M_{\rm Ni}}{M_{\rm \odot}}\,=\,0.61\,\times\,{\rm log}\frac{E_{\rm K}}{\rm foe}\,-1.37,
\end{equation}
which is flatter than the corresponding log\,$L_{\rm prog}$-weighted relation. Moreover, this relation deviates from the predictions of core-collapse simulations, with nearly all objects in the sample lying outside the convex hull of the models. At a fixed log\,$E_{\rm K}$, the explosion models largely overestimate log\,$M_{\rm Ni}$. 

This systematic discrepancy is a common outcome when using wind models or scaling relations. For comparison, we plot the log\,$E_{\rm K}$–log\,$M_{\rm Ni}$ relations derived in \citet{martinez22b,martinez22c} and \citet{muller_t17}. In the former, parameters are inferred from light curve fits using wind models, whereas in the latter, a global fitting approach developed by \citet{pejcha15a,pejcha15b} is applied, with subsequent calibration using the scaling relations from \citet{LN85} and \citet{popov93}. These comparison samples populate a similar region as our wind model results, consistently lying below the core-collapse simulation predictions. For illustration purposes, only the linear regression results are shown.

As discussed in \S5.1 of \citet{fang25d}, this discrepancy arises naturally when wind models are applied: $E_{\rm K}$ tends to be overestimated, and all objects in the left panel of Figure~\ref{fig:e_ni} shift to the right, moving away from the convex hull of the core-collapse simulations (see also Appendix B).

\section{Conclusion}
In this work, we investigate the mutual relationships among helium core mass ($M_{\rm He\,core}$, proxied by log\,$L_{\rm prog}$), explosion energy ($E_{\rm K}$), and explosive nucleosynthesis yield (represented by $M_{\rm Ni}$) for a well-observed sample of 32 SNe II. The results are compared with the predictions from recent neutrino-driven core-collapse simulations, which can be summarized as follow:

\begin{itemize}
    \item For the observational sample, a moderately strong and statistically significant correlation can be discerned between log\,$L_{\rm prog}$ and $E_{\rm K}$ ($\rho$\,=\,0.59$^{+0.11}_{-0.12}$, $p$\,\( < \) 2.4$^{+16.3}_{-2.2}\,\times\,10^{-3}$), suggesting that RSG with more massive helium core tends to produce more energetic explosion;
    \item In the regime log\,$L_{\rm prog}$\,>\,4.6 (corresponding to $M_{\rm He\,core}$\,>\,2.5\,$M_{\odot}$), the observed log\,$L_{\rm prog}$–$E_{\rm K}$ relation agrees well with predictions from core-collapse simulations. Below this threshold (log\,$L_{\rm prog}$\,<\,4.6), the agreement becomes poorer. In particular, while observed SNe II can have inferred $E_{\rm K}$ as large as 0.5 foe, the models consistently predict $E_{\rm K}\sim$0.1 foe. However, this discrepancy can be mitigated by accounting for uncertainties in log\,$L_{\rm prog}$ estimates and host-galaxy extinction.
    \item In the log\,$E_{\rm K}$–log\,$M_{\rm Ni}$ plane, observed SNe II tend to exhibit slightly lower log\,$M_{\rm Ni}$ than predicted by simulations at a given log\,$E_{\rm K}$. Nevertheless, the majority of the sample (26 out of 32, or over 80\%) falls within the range predicted by simulations when uncertainties are taken into account. Moreover, the slope and scatter level of the observed log\,$E_{\rm K}$–log\,$M_{\rm Ni}$ relation match well with the parametric SN models in \citet{schneider25}.
\end{itemize}

These correlations agree closely with the predictions of recent neutrino-driven core-collapse simulations if nebular spectroscopy is employed as additional constraint. As a comparative study, the sample is also modeled using light curves from RSGs where a specific log\,$L_{\rm prog}$-$M_{\rm Henv}$ relation is imposed, a method commonly adopted in the literature. We show that the inferred values of $E_{\rm K}$ and $M_{\rm Ni}$ can be systematically biased, and hence the above correlations modified. This highlights how assumptions about the hydrogen-rich envelope can significantly affect the inferences about the core’s properties, even though the two regions become largely decoupled in the late stages of stellar evolution. 

The key methodological advancement in our analysis is the inclusion of log\,$L_{\rm prog}$, estimated from nebular spectroscopy, into light curve modeling. This approach allows us to treat $M_{\rm Henv}$ as a free parameter to account for uncertainties in mass-loss mechanisms. Moreover, log\,$L_{\rm prog}$ is used as a proxy for the helium core mass, rather than the more commonly examined relations involving ejecta mass ($M_{\rm ej}$) or $M_{\rm ZAMS}$ (e.g., \citealt{chugai14,pumo17,morozova18,martinez19,martinez22c}). For SNe II, a significant portion of both $M_{\rm ej}$ and $M_{\rm ZAMS}$ resides in the hydrogen-rich envelope, which is sensitive to the uncertain mass-loss rates but decoupled from the core-collapse. Directly examining the core mass-$E_{\rm K}$ relation offers a physically grounded diagnostic of the explosion mechanism. To our knowledge, this represents the first systematic effort to investigate such a correlation in SNe II.

These findings support the conclusion that recent neutrino-driven core-collapse simulations successfully reproduce the key explosion properties of SNe II, and it seems that no additional power source is required (see, e.g., \citealt{sukhbold17,matsumoto25}). We acknowledge that some tensions remain, particularly the large scatter in the log\,$L_{\rm prog}$–$E_{\rm K}$ relation at the low-luminosity (low-core-mass) end. However, this discrepancy is notably reduced compared to earlier works that relied on wind-based $M_{\rm ZAMS}$ estimates; we no longer observe anomalously energetic explosions from low-mass progenitors.

Several caveats of this work should be noted: the progenitor models in this work are one-dimensional, hydrostatic, partially-stripped RSGs with \(T_{\rm eff}\,\sim\,\)3650\,K. Relaxing any of these assumptions could potentially change our conclusions. For example, \S4.3 of \citet{fang25d} shows that cooler RSG progenitors yield lower inferred $E_{\rm K}$. Radial pulsations, commonly observed in RSGs (e.g., \citealt{yang11,soraisam18}), are decoupled from the final core evolution, allowing explosions to occur at any phase of the pulsation cycle. The radius of the pulsating RSG at shock breakout can therefore differ markedly from the hydrostatic case assumed in this work \citep{yoon_pulse,goldberg20_pulse,sengupta25,suzuki25}, which can potentially lead to scatter in inferred $E_{\rm K}$, especially for progenitors with relatively massive $M_{\rm He\,core}$ (log\,$L_{\rm prog}$) that may develop pulsations with large amplitudes \citep{bronner25,laplace25}. Finally, asphericity, whether due to the convective nature of the RSG's envelope \citep{goldberg22a,ma24}, seeded during core collapse \citep{burrows24a,fang24,nagao24} or developing in the subsequent shock propagation \citep{goldberg22b}, can also modify light curve properties \citep{vartanyan24}. It is important to quantify the size of these effects on SNe II parameter inference in future studies.

\software{\texttt{MESA} \citep{paxton11, paxton13, paxton15, paxton18, paxton19, mesa23}; \texttt{STELLA} \citep{blinnikov98,blinnikov00,blinnikov06}; \texttt{SciPy} \citep{scipy}; \texttt{NumPy} \citep{numpy}; \texttt{Astropy} \citep{astropy13,astropy18,astropy22}; \texttt{Matplotlib} \citep{matplotlib}; \texttt{emcee} \citep{emcee}}

\begin{acknowledgements}
The authors are grateful to the referee for comments that helped to improve the manuscript.
QF acknowledges support from the JSPS KAKENHI grant 24KF0080. HN is supported by Grant-in-Aid for Scientific Research (23K03468), the
NINS International Research Exchange Support Program, and the HPCI
System Research Project (Project ID: hp250006, hp250226, hp250166). TJM is supported by the Grants-in-Aid for Scientific Research of the Japan Society for the Promotion of Science (JP24K00682, JP24H01824, JP21H04997, JP24H00002, JP24H00027, JP24K00668) and by the Australian Research Council (ARC) through the ARC's Discovery Projects funding scheme (project DP240101786). SN data used in this work are retrieved from the Open Supernova Catalog (\citealt{open_SN}), the Weizmann Interactive Supernova Data Repository (WISeREP; \citealt{wiserep}) and the UC Berkeley Filippenko Group's Supernova Database (SNDB; \citealt{sndb_origin,SNDB}).
\end{acknowledgements}

\newpage
\begin{appendix}
\section{The effect of host galaxy extinction}
\setcounter{figure}{0}
\renewcommand\thefigure{A\arabic{figure}}
In this Appendix, we re-derive the log\,$L_{\rm prog}$-$E_{\rm K}$-$M_{\rm Ni}$ relations from nebular phase spectroscopy and plateau light curve modeling using only the Galactic foreground extinction $E(B-V)_{\rm MW}$. This removes the large, and often uncertain, host-galaxy reddening, which is typically estimated via empirical relations (Na ID equivalent width or assumed intrinsic colors) that lack firm physical calibration. We make two exceptions: for SN 2013am we adopt $E(B-V)$\,=\,0.45\,mag, otherwise it would be too faint to be modeled by the light curve grid, and for SN 2009dd we use $E(B-V)$\,=\,0.20\,mag in order to match its observed color evolution.

Applying only $E(B-V)_{\rm MW}$, the inferred $E_{\rm K}$ and $M_{\rm Ni}$ both decrease systematically compared to the main‐text values, yet the overall agreement between data and models improves. In the log\,$L_{\rm prog}$-$E_{\rm K}$ plane, all SNe now lie within the model‐predicted range, with the exceptions of SN 2016X and SN 2004et, which appears intrinsically more energetic. Our aim here is not to argue for negligible host extinction, but to quantify how uncertain reddening biases comparisons with core‐collapse models.

Importantly, although the absolute values of $E_{\rm K}$ and $M_{\rm Ni}$ are sensitive to the assumed extinction, their mutual relations are not: reddening shifts both quantities along nearly the same direction. We select three test cases (SNe 2004et, 2012A, and 2014G) and add
$E(B-V)$ values from 0.0 to 0.4 mag to these objects, finding
\[
{\rm log}\,E_{\rm K}\,\propto\,1.33\,E(B-V),
\]
\[
{\rm log}\,M_{\rm Ni}\,\propto\,1.09\,E(B-V),
\]
which imply
\[
    {\rm log}\,M_{\rm Ni}\,\propto\,\,0.94\,\times\,{\rm log}\,E_{\rm K}\,+\,0.16\,\times\,E(B-V).
\]
Compared with Equation~\ref{eq:ek_ni_obs}, applying $E(B-V)$\,=\,0.20\,mag (the $B-V$ color scatter of unreddened SNe II in \citealt{deJaeger18}) leads to 0.032\,dex difference in the estimated log\,$M_{\rm Ni}$, much smaller than the observed scatter.

Indeed, linear regressions in the log\,$E_{\rm K}$-log\,$M_{\rm Ni}$ plane, performed with and without host extinction, yield virtually identical results. Since it is robust against both reddening and distance uncertainties, we propose the log\,$E_{\rm K}$-log\,$M_{\rm Ni}$ relation as an especially powerful observational test of core-collapse explosion physics.

\begin{figure*}
\epsscale{0.8}
\plotone{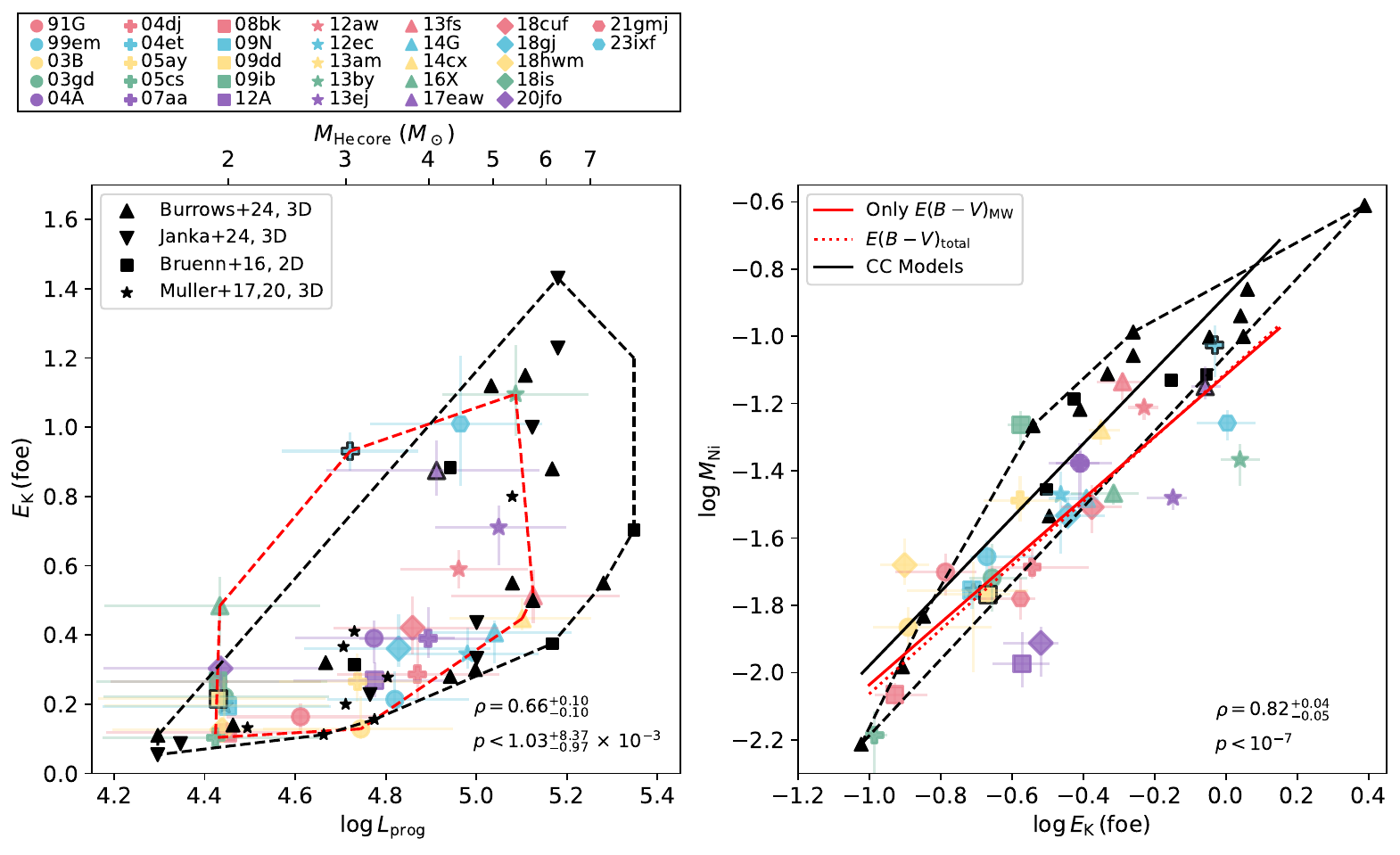}
\centering
\caption{Comparison between log\,$L_{\rm prog}$-$E_{\rm K}$ and log\,$E_{\rm K}$-log\,$M_{\rm Ni}$ for SNe II in the sample. Same as the corresponding figures in the main text except that the observed light curves are only corrected for $E(B-V)_{\rm MW}$.}
\label{fig:unreddened}
\end{figure*}

\newpage
\section{Degeneracy in light curve modeling}
\setcounter{figure}{0}
\renewcommand\thefigure{B\arabic{figure}}
In this Appendix, we will explain in more details why enforcing the log\,$L_{\rm prog}$-$M_{\rm Henv}$ relation for the progenitor models by assuming a specific stellar wind model (for example, the \texttt{KEPLER} models which applied the mass-loss rates from \citealt{wind}) will lead to overestimation of $E_{\rm K}$ in light curve modeling. 

In \citet{fang25a}, the scaling relations between the plateau magnitude ($V_{\rm p}$) and duration ($t_{\rm p}$) are given as
\[
-V_{\rm p}\,\sim\,-\,0.96\,{\rm log}\,M_{\rm Henv}\,+\,2.03\,{\rm log}\,E_{\rm K}\,+\,1.28\,{\rm log}\,R_{\rm prog}
\]
\[
{\rm log}\,t_{\rm p}\,\sim\,0.21\,{\rm log}\,M_{\rm Ni}\,+\,0.55\,{\rm log}\,M_{\rm Henv}\,-0.31\,{\rm log}\,E_{\rm K}\,-\,0.13\,{\rm log}\,R_{\rm prog}.
\]
Similar scaling relations are also found in \citet{goldberg19}. 

\begin{figure}
\epsscale{0.4}
\plotone{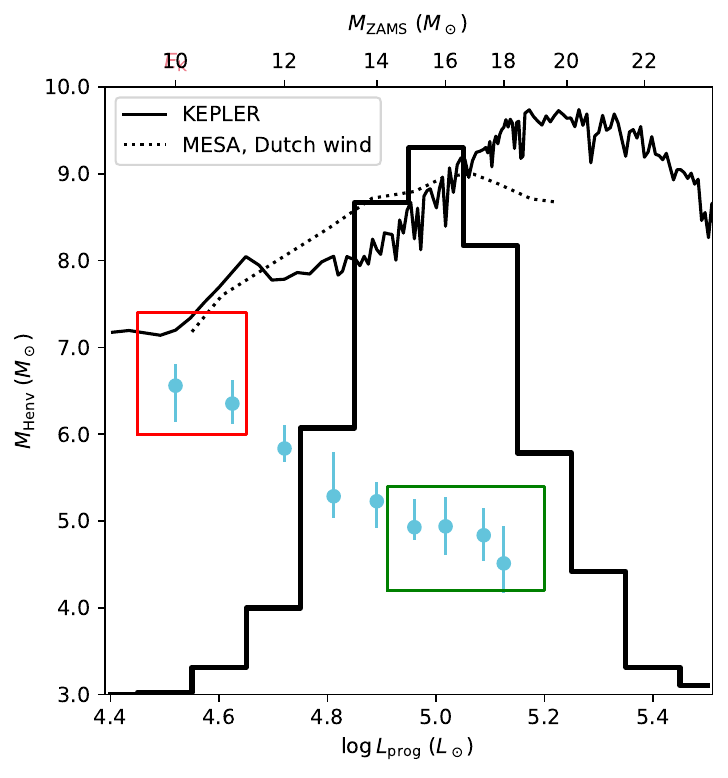}
\centering
\caption{The dependence of inferred $M_{\rm Henv}$ for SN 2014G from light curve modeling based on fixed log\,$L_{\rm prog}$ (light blue scatter points). The black histogram is log\,$L_{\rm prog}$ independently estimated from nebular spectroscopy. The black and dotted solid lines are the log\,$L_{\rm prog}$-$M_{\rm Henv}$ relations of \texttt{KEPLER} models \citep{kepler16} and \texttt{MESA} models with \texttt{Dutch} wind (wind efficiency $\eta=1$. Other parameters, such as mixing length and overshooting, are the same as the ones in the main text). The green box marks $M_{\rm Henv}$ with additional constraint from nebular spectroscopy, while the red box marks values favored by the wind models.}
\label{fig:Henv}
\end{figure}

For fixed $M_{\rm Ni}$ (which can be constrained from the radioactive decay tail) and log\,$L_{\rm prog}$ (which is equivalent to $R_{\rm prog}$ at fixed $T_{\rm eff}$), $E_{\rm K}$ sets the plateau magnitude. Although strictly speaking, $M_{\rm Henv}$ can also play a role, but its effect is weaker than $E_{\rm K}$ given the smaller coefficient and parameter range (see Figure~10 of \citealt{fang25a}). Consequently, progenitor models with smaller log\,$L_{\rm prog}$ ($R_{\rm prog}$) require larger $E_{\rm K}$ to reproduce the same observed plateau magnitude. Once $E_{\rm K}$ is determined, $M_{\rm Henv}$ can then be inferred from $t_{\rm p}$. Thus, progenitor models with smaller log\,$L_{\rm prog}$, therefore larger $E_{\rm K}$, naturally imply larger $M_{\rm Henv}$ according to the scaling relation of $t_{\rm p}$. These arguments qualitatively explain the degeneracy patterns of $E_{\rm K}$ (Figure~\ref{fig:weight}) and $M_{\rm Henv}$ (Figure~\ref{fig:Henv}). 

In \citet{fang25c} and in this work, additional constraint from nebular spectroscopy restricts $M_{\rm Henv}$ to the values within the green box in Figure~\ref{fig:Henv}. By contrast, stellar evolution models such as \texttt{KEPLER} or \texttt{MESA} with the \texttt{Dutch} wind scheme predict log\,$L_{\rm prog}$-$M_{\rm Henv}$ relations trend in an opposite direction against the degeneracy curve, as shown in Figure~\ref{fig:Henv}. If these relations are imposed on progenitor models for light curve modeling, the fitting converges toward the red box in Figure~\ref{fig:Henv}, minimizing the mismatch between the degeneracy curve and the stellar evolution predictions. As a result, if an SN II originate from massive star suffer from enhanced mass-loss, which is argued to be common in \citet{fang25a,fang25c}, enforcing a specific log\,$L_{\rm prog}$-$M_{\rm Henv}$ relation from these stellar evolution models will inevitably underestimate log\,$L_{\rm prog}$ and thus overestimate $E_{\rm K}$ according to the scaling relations. This explains why wind-based progenitor models often yield cases with low log\,$L_{\rm prog}$ but anomalously high $E_{\rm K}$ (right panel of Figure~\ref{fig:mass_e}).

\end{appendix}

{}
\end{document}